\documentclass[10pt,a4paper]{article}
\usepackage{jheppub_kim}
\usepackage{pdflscape}
\usepackage{amsmath}
\usepackage{amssymb}
\usepackage{dcolumn}
\usepackage{bm}
\usepackage{color}
\usepackage{epsfig}
\usepackage{amsfonts}
\usepackage{amsmath}
\usepackage{graphicx}
\usepackage{subfigure}
\usepackage{dcolumn}
\UseRawInputEncoding

\begin{document}
\title{Relativistic model of anisotropic star with Bose-Einstein density depiction in $f(T)$ gravity}

\author[a,b]{Samprity Das}
\author[c]{Prabir Rudra}
\author[d]{Surajit Chattopadhyay}

\affiliation[a]{Department of Mathematics, Amity University, Kolkata, New Town, Rajarhat, Kolkata 700135, India}

\affiliation[b]{Department of Mathematics, Shibpur Dinobundhoo Institution (College), Howrah 711102, West Bengal, India}

\affiliation[c] {Department of Mathematics, Asutosh College,
Kolkata-700 026, India}

\affiliation[d] {Department of Mathematics, Amity University, Rajarhat, New Town, Kolkata 700135, India.}

\emailAdd{samprity.das@s.amity.edu}
\emailAdd{prudra.math@gmail.com}
\emailAdd{schattopadhyay1@kol.amity.edu}

\abstract{This article presents a new model for anisotropic compact stars that are confined to physical dark matter in the background of $f(T)$ teleparallel gravity. The model is based on the equation of state (EoS) of the bag model type and the Bose-Einstein dark matter density profile. The derived solutions meet the energy conditions, the causality conditions, and the required conditions on the stability factor and adiabatic index, indicating that they are physically well-behaved and represent the physical and stable matter configuration. We also determine the maximum mass, surface redshift, and compactness parameter at the surface. Interestingly, all of these numbers fall within the specified range, supporting the physical viability of our proposal. Additionally, the various masses that are derived for varying the model parameter $k$ correspond to five compact, realistic compact objects, including LMC X-4, Her X-1, 4U 1538-52, SAX J1808.4-3658, and Cen X-3. We have also illustrated the radially symmetric profiles of energy density and the moment of inertia for non-rotating stars.}

\maketitle

\section{Introduction}
At both cosmological and astrophysical scales, Einstein's general relativity (GR) theory has been shown to be very successful in explaining gravitational phenomena. This theory of gravity reaches significant accomplishments through the intrinsic gravitational systems, such as star and galaxy structures, which comprise a significant fraction of our observable Universe. In addition to other unknown facts, understanding the development of these self-gravitating systems is essential to learning about the universe's age, composition, and evolution. One of the most important phases in the evolution of celestial structures is the process of gravitational collapse leading to a celestial structure's demise, which produces compact star objects. Compact star objects, or endpoints in the evolution of common astronomical formations, are great places to study the properties and composition of high-density matter. Recently, many compact stellar objects with high densities have been discovered \cite{str1}, and often, these objects are confused for pulsars, which are rotating stars with strong magnetic fields.

Stars are scattered throughout gas and dust clouds containing unevenly distributed components in the majority of galaxies. All active stars eventually reach a stage in their development when the internal nuclear fusion radiation pressure is not enough to survive space's gravitational pull. At this time, the star collapses under the weight of itself, resulting in stellar death \cite{str2}. For the majority of stars, this is the process that yields a compact star, also referred to as an incredibly dense and compact stellar remnant. Put another way, compact stars—which also include white dwarfs, neutron stars, black holes, and quark stars—mark the end of the evolutionary chain of common stars. Phase separations in the early cosmos following the Big Bang might have an impact on compact star formation. Compact stars are characterized by an exceptionally high density and the lack of nuclear activities within them. This prevents them from being able to sustain themselves against gravity. In neutron stars and white dwarfs, the pressure of deteriorated gas acts as a deterrent to gravity. In black holes, the star material is infinitely compressed because of the complete dominance of gravity over all other forces, resulting in infinite density \cite{str3}. When a sun-like star completely collapses owing to nuclear fuel exhaustion, a dense white dwarf is produced. On the other hand, when the inert iron cores of stars with masses equal to or greater than ten times that of the Sun collapse, an extraordinarily dense neutron star or black hole is produced. The smallest red dwarfs haven't died yet, but according to stellar models, they will eventually become low-mass white dwarfs as they finally run out of hydrogen fuel. Until then, they will progressively get brighter and hotter.

When solving problems involving compact stars in General Relativity (GR), it is customary to assume that objects have spherically symmetric and isotropic characteristics. It is not always required to have perfect isotropy and homogeneity in order to characterize the overall physical properties of compact star objects in astrophysics, even though they may have features that can be solved. Since the presence of separate components of radial pressure $(p_{r})$ and tangential pressure $(p_{t})$ cause inhomogeneity, the fluid pressure may therefore consist of these two distinct components that provide an anisotropic factor $(\Delta=p_{t}-p_{r})$. This could impede the internal system of matter distribution from having an idealized isotropic scenario. This idea was first brought up by Ruderman \cite{be1} in his extensive analysis of the behavior and structure of pulsars.
Subsequently, several scientists raised the issue in their papers \cite{be2, be3, be4,bib18}. Different condensate states (such as pion condensates, meson condensates, etc.), superfluid 3A, a mixture of various fluid types, rotational motion, the presence of a magnetic field, phase transition, relativistic particles in compact stars, etc. are all thought to contribute to this anisotropy in the very high-density region of the core \cite{be5, be6, be7, be8, be9, be10,bib17}. Nevertheless, gravitational tidal forces represent another possible contributing factor to anisotropy in compact stars. This is thought to be the reason for the anisotropy of stars' fluid distribution and the ensuing deformation \cite{be11, be12, be13, be14, be15, be16}. In this work, we stress on the anisotropic nature of compact stars.

Dark matter (DM) is a form of hypothetical material that makes up about $27\%$ of the universe's mass-to-energy ratio. Although many candidates, including axions and wimps, are known from supersymmetric string theory and particle physics, there hasn't been a direct scientific detection of dark matter yet. Nevertheless, several astrophysical events, including the formation of galactic rotation curves \cite{dm1}, the dynamics of galaxy clusters \cite{dm2}, and the cosmological scales of anisotropies found by PLANCK in the cosmic microwave background \cite{dm3}, seem to be influenced by dark matter, according to recent experimental data. Dark matter is one of the biggest scientific mysteries of the modern era. There are numerous hypotheses for its identity because so little is known about its microphysical properties.

The quantum statistics of integer spin particles, or bosons, have been a key area of research in theoretical and experimental physics since Bose's conception \cite{bec1} and Einstein's generalization \cite{bec2, bec3}. The bosonic systems' phase change to a condensed state is their most significant feature. This type of quantum bosonic system is known as a Bose-Einstein Condensate (BEC) because all of the particles are in the same quantum ground state. In both coordinate and momentum space, a system of this kind is physically typified by a strong peak over a wider distribution. The particles within a BEC exhibit mutual correlation and wavelength overlap. It is believed that many fundamental processes in condensed matter physics are understood in large part because of the Bose-Einstein condensation process. For instance, assuming a Bose-Einstein condensation mechanism one can explain the superfluidity of low-temperature liquids, such as $^{3} He$ \cite{bec4}. The notion that Bose-Einstein condensation may also occur in bosonic systems occurring at the astrophysical or cosmological scales cannot be immediately ruled out because it is a phenomenon that has been observed and extensively investigated in Earthly systems. Dark matter is thought to be a cold bosonic gravitationally bounded system and is necessary to explain the dynamics of the neutral hydrogen clouds at great distances from the galaxy centers. Thus, dark matter could also exist as a Bose-Einstein condensate \cite{bec5, bec6, bec7}. A systematic investigation into the characteristics of Bose-Einstein Condensed galactic dark matter halos can be found in \cite{bec8}. Recent studies have looked closely at the cosmological and astrophysical consequences of Bose-Einstein Condensed dark matter \cite{bec9, bec10, bec11, bec12, bec13, bec14, bec15, bec16}. Furthermore, it has long been thought that some types of BEC may exist in neutron stars \cite{bec17}. Driven by the aforementioned considerations, we now introduce a novel framework for anisotropic compact stars limited to physical deep space, which is based on the Bose-Einstein condensate DM density profile and features an equation of state akin to the bag model.

Numerous suggestions for various types of modified gravity have been applied over time to address a wide range of physical issues \cite{ft1, ft2, ft3, ft4, ft5, ft6,bib16,bib19}. It is important to note that Riemann geometry, the foundation of GR, is based on curvature that is obtained from metric. This occurred primarily because, at the time that GR was being proposed, it was the only geometric tool that was sufficiently developed \cite{ft7}. Other geometric techniques, including torsional models that are embodied in teleparallel gravity (TG), have been appropriately developed throughout the ensuing decades \cite{ft8, ft9, ft10, ft11}. TG has gained enormous
popularity in recent years and relies on the exchange of the curvature-based Levi-Civita connection with the
torsion-based teleparallel connection. When thinking about novel physics at the level of gravitational theory, this image might offer a more sensible method. In addition to satisfying metricity and being curvature-free, the teleparallel connection forms the foundation of an innovative gravitational framework \cite{ft12}. Nevertheless, there is a certain set of gravitational contractions that result in the teleparallel equivalent of general relativity (TEGR), which is, at the level of the classical field equations, dynamically equal to GR (see \cite{ft13} for a discussion). Accordingly, when it comes to astrophysical and cosmic phenomenology, both GR and TEGR have the same predictions; but, when taking into account IR completions, there might be discrepancies \cite{ft14}. In this case, the action in the torsion scalar $T$ is linear. The same reasoning that applies to conventional modified gravity in the GR regime can now be used to investigate alterations to TG through the TEGR action. $f(T)$ teleparallel gravity, which generates generically second-order equations of motion for all spacetimes based on an arbitrary function of the torsion scalar, is one well-known method for generalizing TEGR \cite{ft15, ft16, ft17, ft18, ft19, ft20, ft21}. Additional extensions have been made to this in several directions, including the use of a scalar field \cite{ft22, ft23, ft24} and additional scalar alterations \cite{ft25, ft26}, such as regular curvature-based modified gravity.

In this work, we are interested in exploring a new model for an anisotropic compact star confined to physical dark matter based on a Bose-Einstein condensate dark matter density profile with a bag model type equation of state in the background of $f(T)$ teleparallel gravity. The combination of BEC in a torsion-based gravity is expected to produce really interesting results. The paper is organized as below: In section 2 we discuss the basic equations of $f(T)$ gravity. In section 3, we find the solutions of the field equations that describe the stellar model in the background of $f(T)$ gravity. Section 4 is dedicated to matching the internal and external solutions at the boundary. In section 5 we perform a physical analysis of the model. A comprehensive equilibrium and stability analysis is performed in section 6. In section 7 we analyze the moment of inertia of the system and finally, the paper ends with some discussion and conclusion in section 8.

\section{Basic equations of $f(T)$ gravity}

The metric tensor, which is defined on a pseudo-Riemannian manifold, serves as the dynamical variable used to construct general relativity. Nonetheless, the tetrad field is employed as a dynamical variable in $f(T)$ gravity, forming an orthonormal basis in tangent space. Let's now indicate Greek indices $(\mu, \nu, ....)$ for space-time coordinates and Latin indices $(i,j,...)$ for tangent space coordinates. Thus $e_i=\partial_i$ and $e^i = dx^i$ denotes the basis vectors and covectors for the tangent space respectively. Also, $e_\mu=\partial_\mu$ and $e^ \mu = dx^ \mu$ denotes the basis vectors and covectors for the space-time coordinates respectively. The relation between bases can be represented by 
\begin{equation}
    \begin{array}{ccc}
    \partial _i=e_i^{\mu }\partial _{\mu } & and &  {dx}^i=e_{\mu }^i {dx}^{\mu } ,
\end{array}
\label{E01}
\end{equation}
and conversely
\begin{equation}
    \begin{array}{ccc}
   \partial _{\mu }=e_{\mu }^i\partial _i & and &  {dx}^{\mu }=e_i^{\mu } {dx}^i ,
\end{array}
\label{E02}
\end{equation}
where the tetrad field $e_{\mu }^i$ and the inverse $e_i^{\mu }$ satisfy the relation $e_j^{\mu } e_{\mu }^i=\delta_j^i$ and $e_j^{\mu } e_{\nu }^j=\delta_\nu ^ \mu$. The relation between the tetrad field and the space-time metric defined as 
\begin{equation}
g_{\mu \nu }=\eta _{{ij}}e_{\mu }^i e_{\nu }^j ,
    \label{E03}
\end{equation}
for $\eta_{ij}=diag[1,-1,-1,-1]$ and $\sqrt{-g}=det[e_{\mu }^i]=e$. With the help of Eq. (\ref{E01}) the tetrad field equations converted as
\begin{equation}
{ds}^2=g_{\mu \nu }{dx}^{\mu }{dx}^{\nu }=\eta _{{ij}}e_{\mu }^ie_{\nu }^j{dx}^{\mu }{dx}^{\nu } .
    \label{E04}
\end{equation}
Now Weitzenböck connection, which is known as curvature-less connection in $f(T)$ gravity theory, given by \cite{bib2}
\begin{equation}
\Gamma _{\mu \nu }^{\alpha }=e_i^{\alpha }\partial _{\nu }e_{\mu }^i=-e_i^{\mu }\partial _{\nu }e_i^{\alpha } .
    \label{E05}
\end{equation}
For  Weitzenböck connection the torsion tensor converted to
\begin{equation}
T_{\mu \nu }^{\alpha }=\Gamma _{\mu \nu }^{\alpha }-\Gamma _{\nu \mu }^{\alpha }=e_i^{\alpha }\left(\partial _{\mu }e_{\nu }^i-\partial _{\nu }e_{\mu }^i\right)
    \label{E06}
\end{equation}
Hereafter the con-torsion and the superpotential tensor can be evaluated respectively by calculating torsion scalar
\begin{equation}
K_{\alpha }^{\mu \nu }=\frac{-1}{2}\left(T_{\alpha }^{\mu \nu }-T_{\alpha }^{\nu \mu }-T_{\alpha }^{\mu \nu }\right) ,
    \label{E07}
\end{equation}
\begin{equation}
S_{\alpha }^{\mu \nu }=\frac{1}{2}\left(K_{\alpha }^{\mu \nu }-\delta _{\alpha }^{\mu }T_{\gamma }^{\gamma \nu }-\delta _{\alpha }^{\nu }T_{\gamma }^{\gamma \mu }\right) .
    \label{E08}
\end{equation}
Here the torsion scalar represented as
\begin{equation}
T=T_{\mu \nu }^{\alpha }S_{\alpha }^{\mu \nu } .
    \label{E09}
\end{equation}
Similar to $f(R)$ gravity, in $f(T)$ gravity the modified gravitational action can be written in terms of torsion scalar in the units of $(G=c=1)$ as
\begin{equation}
S\left[e_{\mu }^i,\phi _A\right]=\int d^4x e \left[\frac{1}{16 \pi }f(T)+\mathcal{L}_{ {matter}}\left(\phi _A\right)\right] ,
    \label{E10}
\end{equation}
where $f(T)$ is function of torsion scalar and $\mathcal{L}_{ {matter}}$ is the Lagrangian density of matter field and $e=\sqrt{-g}=det[e_{\mu }^i]$. 
Now Eq. (\ref{E10}) gives the field equation for $f(T)$ gravity \cite{bib3,bib4}
\begin{equation}
    S_{\mu }^{\nu \lambda }\partial _{\lambda }T f_{{TT}}+\left[e^{-1}e_{\mu }^i \partial _{\lambda }\left({ee}_i^{\alpha } S_{\alpha }^{\nu \lambda }\right)+ T_{\theta \mu }^{\alpha } S_{\alpha }^{\nu \theta }\right] f_T + \frac{1}{4}\delta _{\mu }^{\nu }f=4 \pi  \tau _{\mu }^{\nu } ,
\label{E11}
\end{equation}
where $\tau _{\mu }^{\nu } $ is the energy moment tensor of the matter. $f_T$ and $f_{TT}$ is first and second order derivative of $f(T)$ with respect to $T$ respectively. For anisotropic fluid distribution the energy moment tensor can be described as
\begin{equation}
T_{\mu }^{\nu }=\left(\rho +p_t\right)u_{\mu }u^{\nu }-p_t\delta _{\mu }^{\nu }+\left(p_r-p_t\right)v_{\mu }v^{\nu } ,
    \label{E12}
\end{equation}
where $u^{\nu }$ and $v^{\nu }$ are four velocity and radial four vectors respectively. $\rho$, $p_r$, and $p_t$ are the density, radial pressure, and tangential pressure respectively. 

\section{Solution of field equations}

For anisotropic stars we assume spherically symmetric metric
\begin{equation}
{ds}^2=e^{\zeta (r)}{dt}^2-e^{\eta (r)}{dr}^2-r^2\left({d\theta }^2+{Sin}^2\theta  {d\phi }^2\right) ,
    \label{E13}
\end{equation}
where $\zeta (r)$ and $\eta (r)$ are dimensionless non-linear functions of radial coordinate. The tetrad matrix from Eq. (\ref{E13}) can be evaluated as
\begin{equation}
e_{\mu }^i={diag}\left[e^{\frac{\zeta (r)}{2}},e^{\frac{\eta (r)}{2}},r,r {Sin}(\theta )\right] ,
    \label{E14}
\end{equation}
for $e=\det \left[e_{\mu }^i\right]=e^{\frac{\zeta (r)+\eta (r)}{2}}r^2{Sin}(\theta )$. Now the torsion term is determined as
\begin{equation}
T(r)=\frac{2 e^{-\eta (r)}}{r}\left(\zeta '+\frac{1}{r}\right) .
    \label{E15}
\end{equation}
Now using Eq. (\ref{E11}) for the metric  Eq. (\ref{E13}) one can derive the the set of equations for $f(T)$ gravity for anisotropic fluid
\begin{equation}
4\pi  \rho =\frac{f}{4}-\frac{f_T}{2}\left(T-\frac{ e^{-\eta (r)}}{r}(\zeta '+\eta ')-\frac{1}{r^2}\right) ,
    \label{E16}
\end{equation}
\begin{equation}
4\pi  p_r=\frac{f_T}{2}\left(T-\frac{1}{r^2}\right)-\frac{1}{4} ,
    \label{E17}
\end{equation}
\begin{equation}
4\pi  p_t=\left(\frac{T}{2}+ e^{-\eta (r)}\left(\frac{\zeta \texttt{"}}{2}+\left(\frac{\zeta '}{4}+\frac{1}{2 r}\right)(\zeta '-\eta ')\right)\right)\frac{f_T }{2} ,
    \label{E18}
\end{equation}
\begin{equation}
\frac{{Cot} \theta }{2 r^2}T' f_{{TT}}=0 .
    \label{E19}
\end{equation}
If and only if $f_{TT}=0$, the exact solution derived from the exterior solution of field equations is the Schwarzschild (anti-) de Sitter solution. As a result, $f_{TT}=0$ must be assumed in order to determine the functional form of $f (T)$ and the solution for self-gravitating compact star models.
Eq. (\ref{E19}) provides the solution as 
\begin{equation}
f=\alpha T+ \beta ,
    \label{E20}
\end{equation}
where $\alpha$ and $\beta$ are arbitrary constants.

Numerous investigations have been conducted to find the proper matter distribution in the interior of compact stars, which also have significant density attributes \cite{bib5,bib6,bib7,bib8}. In this study we have chosen \emph{Bose-Einstein Dark Matter} density profile \cite{bib9} and the corresponding linear EoS to study the physical attributes of the compact stars in the background of $f(T)$ gravity.
\begin{equation}
\rho =\frac{a}{k r}{Sin}(k r) ,
    \label{E21}
\end{equation}
\begin{equation}
{pr}=\eta _1  \rho -\eta _2 ,
    \label{E22}
\end{equation}
where $a (km^{-1})$, $k$, $\eta _1$, and $\eta _2 (km^{-2})$ are non-zero positive constants. 
Hence the mass of the matter distribution can be evaluated from density profile using Eq. (\ref{E21})
\begin{equation}
m=\int _0^r4 \pi  x\frac{a}{k }{Sin}(k x)dx
    \label{E23}
\end{equation}
which gives
\begin{equation}
m=\frac{4 a \pi  (-k r {Cos}(k r)+{Sin}(k r))}{k^3}
    \label{E24}
\end{equation}
Since Bose-Einstein condensation has been observed and studied in great detail in Earthly systems, it cannot be immediately ruled out that it may also occur in bosonic systems happening at the astrophysical or cosmic scales. It is believed that dark matter is a cold bosonic gravitationally constrained system, which is required to explain the neutral hydrogen cloud dynamics at large distances from the galactic centers. Consequently, it is possible that dark matter exists as a Bose-Einstein condensate \cite{bec5, bec6, bec7}. \cite{bec8} provides a systematic study of the properties of Bose-Einstein Condensed galactic dark matter halos. 

\section{Matching of the internal and external solutions at the
boundary}

When considering compact objects in the GR backdrop and weighing the different matching requirements, the Schwarzschild solution is often regarded as the optimal choice. The Jebsen-Birkhoff theorem states that the solution to the Einstein field equations (EFE) must be asymptotically flat and static given spherically symmetric spacetime. When we encounter modified TOV equations with zero pressure and energy density, the exterior geometry solution in modified gravitational theories might be different from the Schwarzschild solution \cite{bib10,bib11}.
\begin{equation}
{ds}^2=\left(1-\frac{2 m}{r}\right){dt}^2-\left(1-\frac{2 m}{r}\right)^{-1}{dr}^2-r^2\left({d\theta }^2+{Sin}^2\theta  {d\phi }^2\right) .
    \label{E25}
\end{equation}
Hence comparing Eq. (\ref{E13}) and Eq. (\ref{E23}) we get
\begin{equation}
e^{\eta }=\left(1-\frac{2 m}{r}\right)^{-1}=\left(1-\frac{8 a \pi  (-k r {Cos}(k r)+{Sin}(k r))}{k^3 r} \right) ^{-1}
    \label{E26}
\end{equation}
\begin{equation}
e^{\zeta }=1-\frac{2 m}{r}=1+\frac{8 a \pi  (k r {Cos}(k r)-{Sin}(k r))}{k^3 r}
    \label{E27}
\end{equation}
From Eq. (\ref{E26}) and Eq. (\ref{E27}) we evaluate the metric potential functions as
\begin{equation}
\zeta '=\frac{-\frac{8 a \pi  (k r{Cos}(k r)-{Sin}(k r))}{k^3 r^2}-\frac{8 a \pi  {Sin}(k r)}{k}}{1+\frac{8 a \pi  (k r {Cos}(k r)-{Sin}(k r))}{k^3 r}}
    \label{E28}
\end{equation}
\begin{equation}
\eta '=-\frac{-\frac{8 a \pi  (k r {Cos}(k r)-{Sin}(k r))}{k^3 r^2}-\frac{8 a \pi  {Sin}(k r)}{k}}{1+\frac{8 a \pi  (k r {Cos}(k r)-{Sin}(k r))}{k^3 r}}
    \label{E29}
\end{equation}
The graphical demonstration of metric potential functions as follows

\begin{figure}
   \begin{center}
       \subfigure[]{\includegraphics[width=0.4\textwidth]{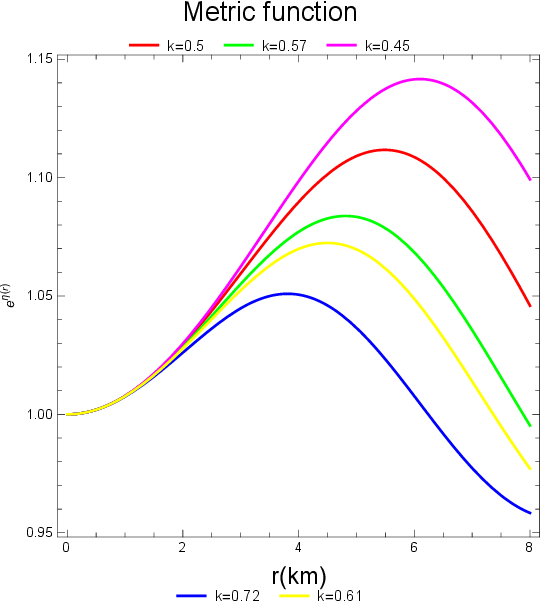}} 
    \subfigure[]{\includegraphics[width=0.4\textwidth]{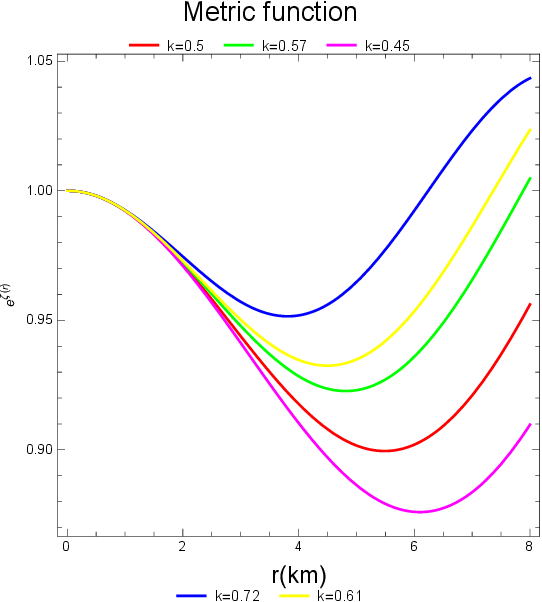}} 
    \caption{(a) Metric potential function from Eq. (\ref{E26}) with respect to $r$ for parametric perturbation $k=0.5, 0.57, 0.45, 0.72, 0.61$ respectively. \\
    (b) Metric potential function from Eq. (\ref{E27}) with respect to $r$ for parametric perturbation $k=0.5, 0.57, 0.45, 0.72, 0.61$ respectively. }
    \label{fig:metric}
   \end{center}
\end{figure}

Fig \ref{fig:metric} shows the evolution of metric potential functions with respect to the radial coordinate. It is evident that in the compact stars, both metric potentials have positive values.

\section{Physical analysis of the model}
In this section, we will perform a physical analysis of the model to check its viability.

\subsection{Density, Pressure, and Anisotropic factor}

In order to analyze physical attributes we evaluated density, radial pressure, and tangential pressure within the compact stars in the background of $f(t)$ gravity theory.
Using Eq. (\ref{E15}) and Eq. (\ref{E19}), $T$ and $f$ have been reconstructed as function of $r$ respectively
\begin{equation}
T(r)=\frac{2 (k-8 a \pi  r {Sin}(k r))}{k r^2}
    \label{E30}
\end{equation}
\begin{equation}
f(r)=\frac{2 \alpha }{r^2}+\beta -\frac{16 a \pi  \alpha {Sin}(k r)}{k r}
    \label{E31}
\end{equation}
Now the density, radial pressure, and tangential pressure of the compact stars evaluated from Eq. (\ref{E16}), Eq. (\ref{E17}), and Eq. (\ref{E18}) 
\begin{equation}
\rho =\frac{\beta}{16 \pi }+\frac{a \alpha {Sin}(k r)}{k r}
    \label{E32}
\end{equation}
\begin{equation}
p_r(r)=-\frac{k r^2-2 k \alpha +32 a \pi  r \alpha {Sin}(k r)}{16 k \pi  r^2}
    \label{E33}
\end{equation}
\begin{multline}
p_t(r)=\frac{1}{8 k^3 \pi  r^3 \left(k^3 r+8 a k \pi  r {Cos}(k r)-8 a \pi {Sin}(k r)\right)} \biggl(\alpha  \biggl(k^6 r^2+  16 a^2 \pi ^2 \left(1+5 k^2 r^2-3 k^4 r^4\right) \\
+8 a \pi \biggl( k^4 r^2 \left(2 -k^2 r^2\right) {Cos}(k r)
 -2 a \pi \left(1+3 k^2 r^2+k^4 r^4\right) {Cos}(2 k r)- k^3 r \left(2+k^2 r^2\right) {Sin}(k r) \\
 -4 a k \pi  r \left(1+k^2 r^2\right) {Sin}(2 k r)\biggr) \biggr) \biggr).
 \label{E34}
\end{multline}
The anisotropic factor $\Delta=p_t-p_r$ has a great impact in the study of physical attributes of compact stars. In the event that $p_t (r) < p_r (r)$, the force will act inward; otherwise, it will act outward [72]. It is observed that when using anisotropic fluid instead of isotropic fluid, the outward force for $p_t (r) > p_r (r)$ aids in the construction of more massive compact objects.
The corresponding anisotropic factor
\begin{multline}
\Delta =\frac{1}{16 k^3 \pi  r^3 \left(k^3 r+8 a k \pi  r {Cos}(k r)-8 a \pi {Sin}(k r)\right)} \biggl( k^6 r^4+32 a^2 \pi ^2 \left(1+k^2 r^2-3 k^4 r^4\right) \alpha +8 a \pi  \biggl( k^4 r^2  \\
\biggl( 2 \alpha + r^2 \left(1-2 k^2 \alpha\right) \biggr) {Cos}(k r)-4 a \pi  \left(1-k^2 r^2+k^4 r^4\right) \alpha {Cos}(2 k r)+k^3 r \left(-2 \alpha+r^2 \left(-1+2 k^2 \alpha\right)\right) {Sin}(k r) \\
+8 a k \pi  r \left(-1+k^2 r^2\right) \alpha {Sin}(2 k r) \biggr) \biggr)
    \label{E35}
\end{multline}
We analyze the first and second order derivatives of the density with respect to the radial coordinate in order to determine the highest possible level of the density in the interior of the compact stars.  
\begin{equation}
\frac{{d\rho }(r)}{{dr}}=\frac{a \alpha {Cos}(k r)}{r}-\frac{a \alpha {Sin}(k r)}{k r^2}
    \label{E36}
\end{equation}
\begin{equation}
\frac{d^2\rho }{{dr}^2}=-\frac{a \alpha \left(2 k r {Cos}(k r)+\left(-2+k^2 r^2\right) {Sin}(k r)\right)}{k r^3}
    \label{E37}
\end{equation}
From Eq. (\ref{E36}) equivalent to zero and for non-zero $a, \alpha$, we get $\frac{tan(kr)}{kr} =1$, which is possible for 
\begin{equation}
\lim_{r\rightarrow 0} \frac{\tan  ({kr})}{{kr}}=1
    \label{E38}
\end{equation}
For Eq. (\ref{E38}), Eq. (\ref{E37}) attains negative value i.e. for $\frac{d\rho}{dr}=0$ and $\frac{d^2 \rho}{dr^2} <0$ we deduce the range of $r$. Hence we can conclude that density attains its maximum level near the core of the compact stars i.e. for $r \rightarrow 0$. 

Now we discuss the graphical representation of the density, radial pressure, tangential pressure, and the corresponding anisotropic factor from Eq. (\ref{E32}), Eq. (\ref{E33}), Eq. (\ref{E34}), and Eq. (\ref{E35})

\begin{figure}
   \begin{center}
       \subfigure[]{\includegraphics[width=0.4\textwidth]{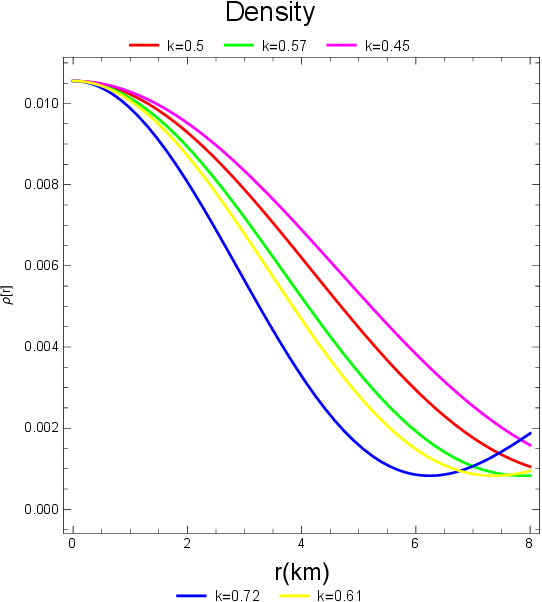}} 
    \subfigure[]{\includegraphics[width=0.4\textwidth]{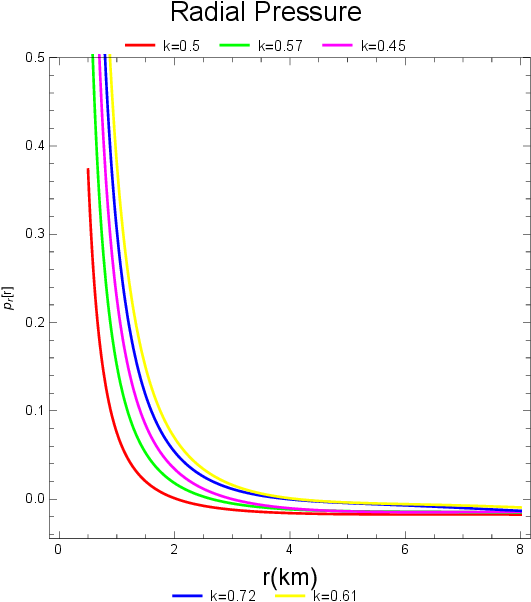}} 
    \caption{(a) Density profile of compact star from Eq. (\ref{E32}) with respect to $r$ for parametric perturbation $k=0.5, 0.57, 0.45, 0.72, 0.61$ respectively. \\
    (b) Radial pressure of compact stars from Eq. (\ref{E33}) with respect to $r$ for parametric perturbation $k=0.5, 0.57, 0.45, 0.72, 0.61$ respectively. }
    \label{fig:density}
   \end{center}
\end{figure}
\begin{figure}
   \begin{center}
       \subfigure[]{\includegraphics[width=0.4\textwidth]{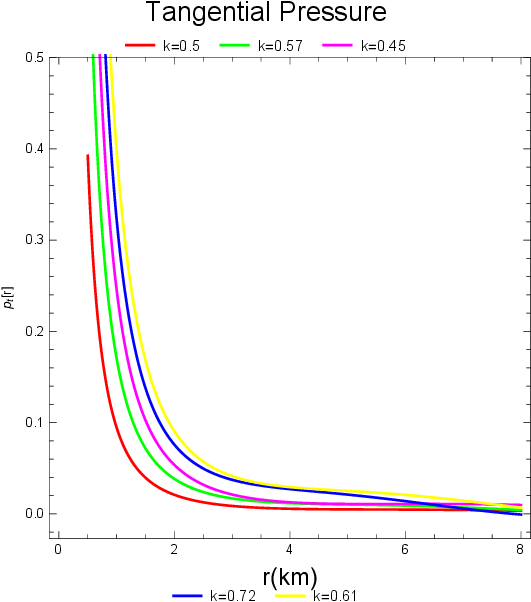}} 
    \subfigure[]{\includegraphics[width=0.4\textwidth]{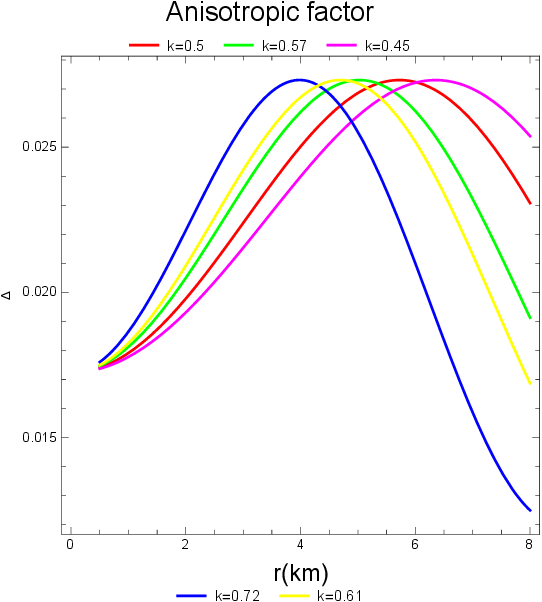}} 
    \caption{(a) Tangential pressure of compact stars from Eq. (\ref{E34}) with respect to $r$ for parametric perturbation $k=0.5, 0.57, 0.45, 0.72, 0.61$ respectively. \\
    (b) Anisotropic factor of compact stars from Eq. (\ref{E35}) with respect to $r$ for parametric perturbation $k=0.5, 0.57, 0.45, 0.72, 0.61$ respectively. }
    \label{fig:ani}
   \end{center}
\end{figure}
The density function for compact stars is shown in the left panel of Fig. \ref{fig:density} with respect to radius. This function ensures that the compact stars reach their highest level close to the core region and then follow a diminishing pattern over the surface. The right panel shows the pressure in the radial direction with respect to the radial coordinate. Here the pressure in the radial direction attains its maximum near the center, has a decaying pattern, and eventually vanishes at the surface of the star. The left panel figure in Fig \ref{fig:ani} demonstrates the pressure in the tangential direction with respect to the radius and shows a decaying pattern near the surface. The right panel figure shows the anisotropic factor throughout the region of the compact star, which stays positive in the interior of the object \cite{bib12}. Also, the anisotropic factor vanishes near the surface, indicating the interior fluid is isotropic in nature. Moreover, for any compact object, $p_{t} > p_{r}$, the anisotropic factor, represented by $\Delta=p_{t}-p_{r}$, indicates that the pressure is not equal in both the transversal and radial directions. The aforementioned data also supports our derived model's absence of singularity.

\begin{figure}
  \begin{center}
       \subfigure[]{\includegraphics[width=0.28\textwidth]{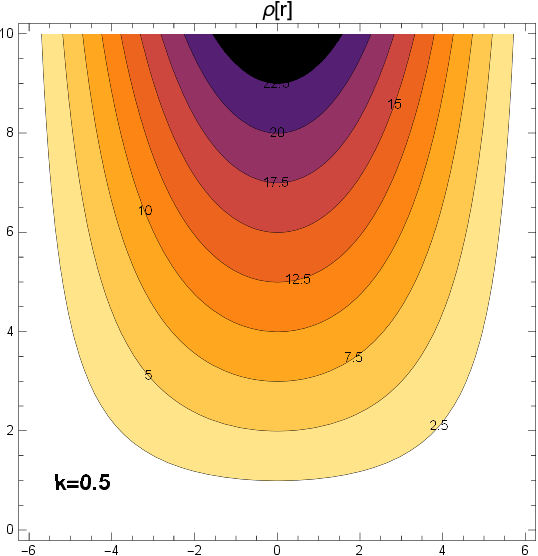}} 
    \subfigure{\includegraphics[width=0.034\textwidth]{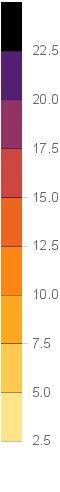}}
    \subfigure[]{\includegraphics[width=0.28\textwidth]{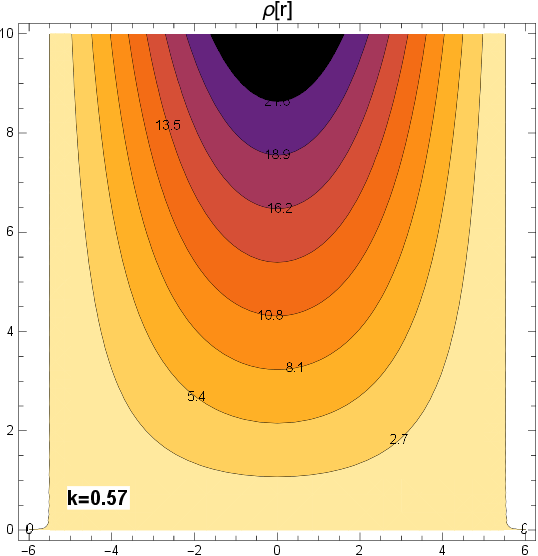}} 
    \subfigure{\includegraphics[width=0.035\textwidth]{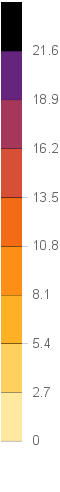}} 
    \subfigure[]{\includegraphics[width=0.28\textwidth]{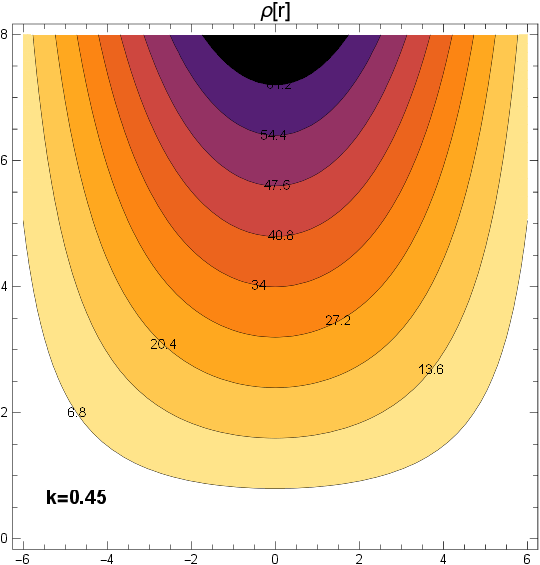}} 
    \subfigure{\includegraphics[width=0.035\textwidth]{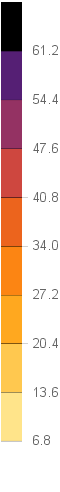}} 
    \caption{(a)Symmetric profile distribution for density of compact star from Eq. (\ref{E32}) with respect to $r$ for parametric perturbation $k=0.5$. \\
    (c) Symmetric profile distribution for density of compact stars from Eq. (\ref{E32}) with respect to $r$ for parametric perturbation $k=0.57$ . \\
    (e) Symmetric profile distribution for density of compact stars from Eq. (\ref{E32}) with respect to $r$ for parametric perturbation $k=0.45$ . }
    \label{fig:contour1}
   \end{center}
\end{figure}
\begin{figure}
  \begin{center}
       \subfigure[]{\includegraphics[width=0.37\textwidth]{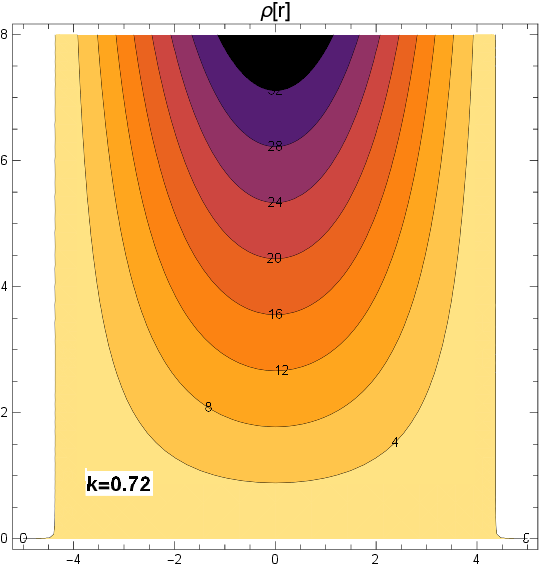}} 
    \subfigure{\includegraphics[width=0.036\textwidth]{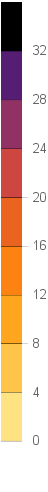}}
    \subfigure[]{\includegraphics[width=0.37\textwidth]{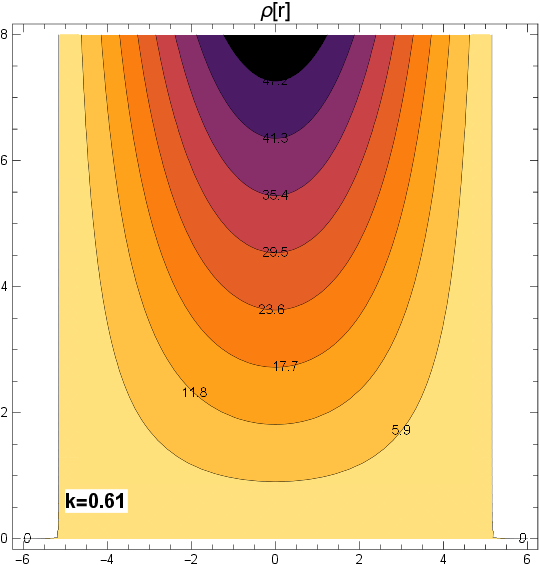}} 
    \subfigure{\includegraphics[width=0.046\textwidth]{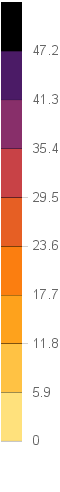}} 
    \caption{(a)Symmetric profile distribution for density of compact star from Eq. (\ref{E32}) with respect to $r$ for parametric perturbation $k=0.72$. \\
    (c) Symmetric profile distribution for density of compact stars from Eq. (\ref{E32}) with respect to $r$ for parametric perturbation $k=0.61$ }. 
    \label{fig:contour2}
   \end{center}
\end{figure}

Now the density gradient, radial pressure gradient, and tangential pressure gradient can be evaluated using Eq. (\ref{E32}),  Eq. (\ref{E33}),  Eq. (\ref{E34}) as follows
\begin{equation}
\frac{{d\rho }(r)}{{dr}}=\frac{a \alpha {Cos}(k r)}{r}-\frac{a \alpha {Sin}(k r)}{k r^2} ,
    \label{E39}
\end{equation}
\begin{equation}
\frac{dp_r(r)}{{dr}}=-\frac{2 k r+32 a k \pi  r \alpha {Cos}(k r)+32 a \pi  \alpha {Sin}(k r)}{16 k \pi  r^2}+\frac{k r^2-2 k \alpha+32 a \pi r \alpha {Sin}(k r)}{8 k \pi  r^3} ,
    \label{E40}
\end{equation}
\begin{multline}
\frac{{dp}_t(r)}{{dr}}=\frac{\alpha}{4 k^3 \pi  r^4 \left(k^3 r+8 a k \pi  r {Cos}(k r)-8 a \pi {Sin}(k r)\right)^2}  \biggl(-96 a^2 k^3 \pi ^2 r-k^9 r^3-144 a^2 k^5 \pi ^2 r^3-64 a^2 k^7 \pi ^2 r^5 \\
-4 a k \pi  r \left(k^6 r^2 \left(7+k^2 r^2\right)+8 a^2 \pi ^2 \left(9+19 k^2 r^2+15 k^4 r^4\right)\right) {Cos}(k r)-16 a^2 k^3 \pi ^2 r \left(-6+3 k^2 r^2+2 k^4 r^4\right) {Cos}(2 k r)+ \\
288 a^3 k \pi ^3 r {Cos}(3 k r)+224 a^3 k^3 \pi ^3 r^3 {Cos}(3 k r)-32 a^3 k^5 \pi ^3 r^5 {Cos}(3 k r)+288 a^3 \pi ^3 {Sin}(k
r)+28 a k^6 \pi  r^2 {Sin}(k r) \\
+576 a^3 k^2 \pi ^3 r^2 {Sin}(k r)+640 a^3 k^4 \pi ^3 r^4 {Sin}(k r)+
4 a k^{10} \pi  r^6 {Sin}(k r)-96 a^3 k^6 \pi ^3 r^6 {Sin}(k r)+192 a^2 k^4 \pi ^2 r^2 {Sin}(2k r) \\
+96 a^2 k^6 \pi ^2 r^4 {Sin}(2k r)+16 a^2 k^8 \pi ^2 r^6 {Sin}(2k r)-96 a^3 \pi ^3 {Sin}(3k r)+192 a^3 k^2 \pi ^3 r^2 {Sin}(3k r)+ 256 a^3 k^4 \pi ^3 r^4 {Sin}(3k r) \\
+32 a^3 k^6 \pi ^3 r^6 {Sin}(3k r)\biggr) .
     \label{E41}
\end{multline}
Graphical representation of Eq. (\ref{E39}), Eq. (\ref{E40}), and Eq. (\ref{E41}) are given in
\begin{figure}
   \begin{center}
       \subfigure[]{\includegraphics[width=0.3\textwidth]{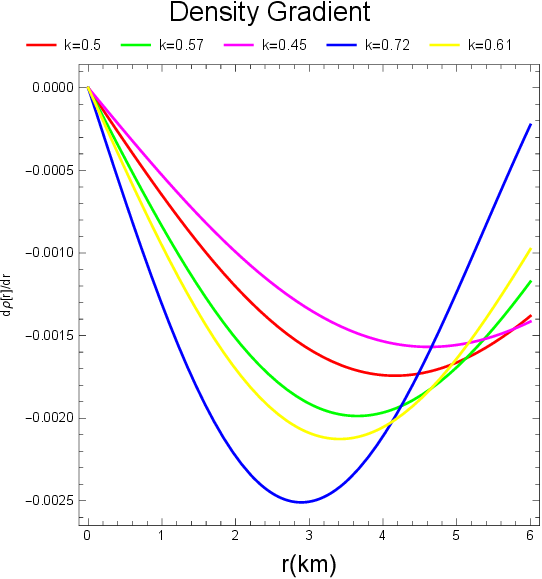}} 
    \subfigure[]{\includegraphics[width=0.3\textwidth]{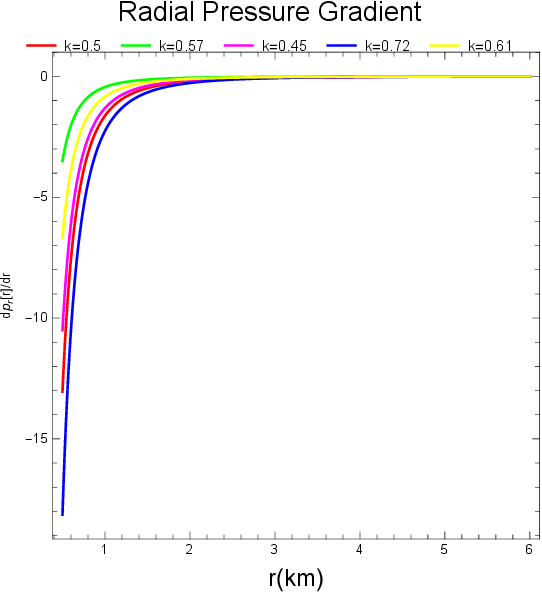}} 
     \subfigure[]{\includegraphics[width=0.3\textwidth]{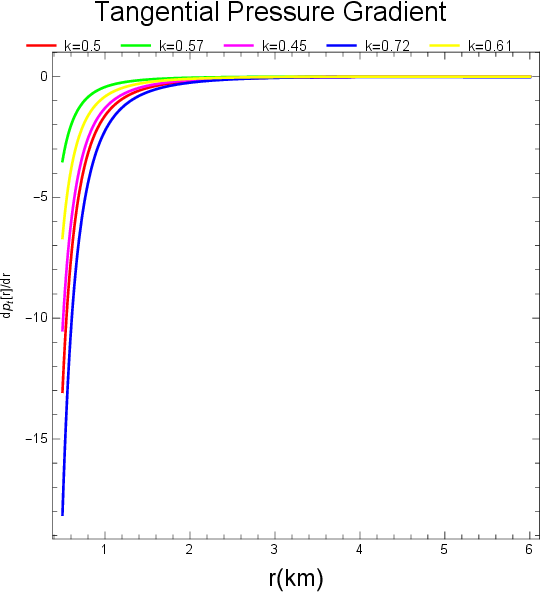}}
    \caption{(a) Density gradient of compact stars from Eq. (\ref{E39}) with respect to $r$ for parametric perturbation $k=0.5, 0.57, 0.45, 0.72, 0.61$ respectively. \\
    (b) Radial pressure gradient of compact stars from Eq. (\ref{E40}) with respect to $r$ for parametric perturbation $k=0.5, 0.57, 0.45, 0.72, 0.61$ respectively. \\
     (c) Tangential pressure gradient of compact stars from Eq. (\ref{E41}) with respect to $r$ for parametric perturbation $k=0.5, 0.57, 0.45, 0.72, 0.61$ respectively.}
    \label{fig:gradient}
   \end{center}
\end{figure}
in Fig \ref{fig:gradient} which demonstrate the first-order derivative of the density, the radial pressure and the tangential pressure with respect to radial coordinates respectively. As depicted in the diagrams the gradient of these physical attributes shows negative trend in the figures. 

\subsection{Energy Conditions}

In this section, we shall discuss the compact object's energy state. When evaluating the matter's unusual behavior inside the compact object, these energy conditions are helpful. \emph{Null Energy Condition (NEC)} and \emph{Strong Energy Condition (SEC)} are the two main categories into which energy states are divided, and their corresponding inequalities are \cite{bib13} \\
$(i)$ $NEC : \rho + p_{r} \geq 0 $ , $\rho + p_{t} \geq 0$ \\
$(ii)$ $SEC : \rho + p_{r} \geq 0 $ , $\rho + p_{t} \geq 0$ , $\rho + p_{r} + 2 p_{t} \geq 0$. \\
The energy requirements are a useful tool for investigating the Hawking-Penrose singularity hypothesis and the second law of black hole thermodynamics \cite{bib13}. Numerous intriguing cosmological occurrences have been explained by energy circumstances \cite{bib14,bib15}. Future singularity studies and our understanding of the DE phase are affected by energy conditions. Moreover, meeting these requirements is essential to understanding the distinctive behavior of matter in the compact structure model. If the energy conditions are satisfied, we can conclude that our star models are feasible.
\begin{figure}
   \begin{center}
       \subfigure[]{\includegraphics[width=0.3\textwidth]{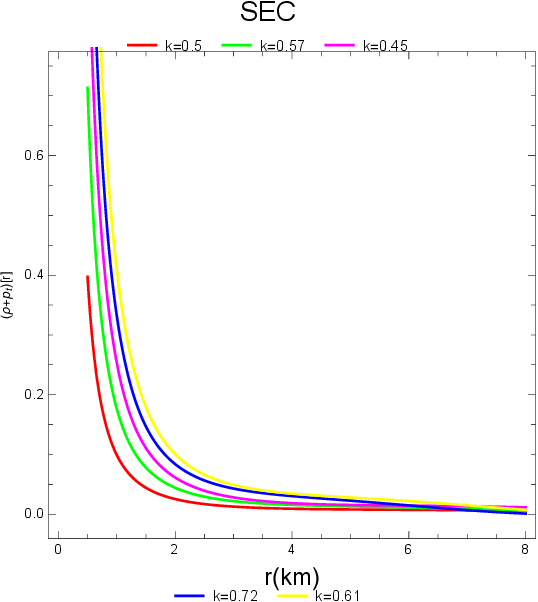}} 
    \subfigure[]{\includegraphics[width=0.3\textwidth]{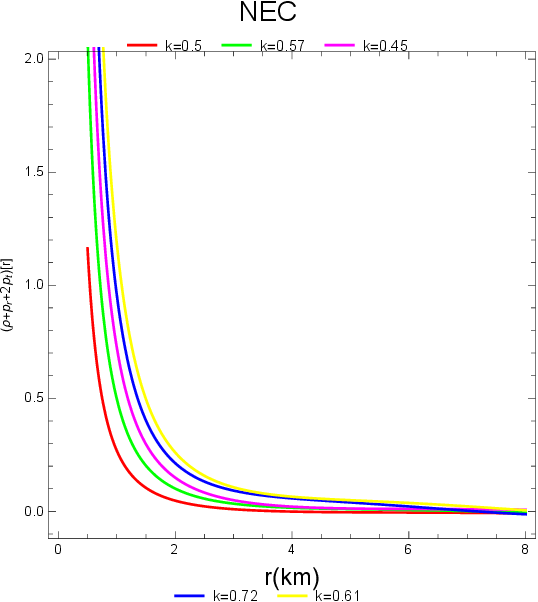}} 
     \subfigure[]{\includegraphics[width=0.3\textwidth]{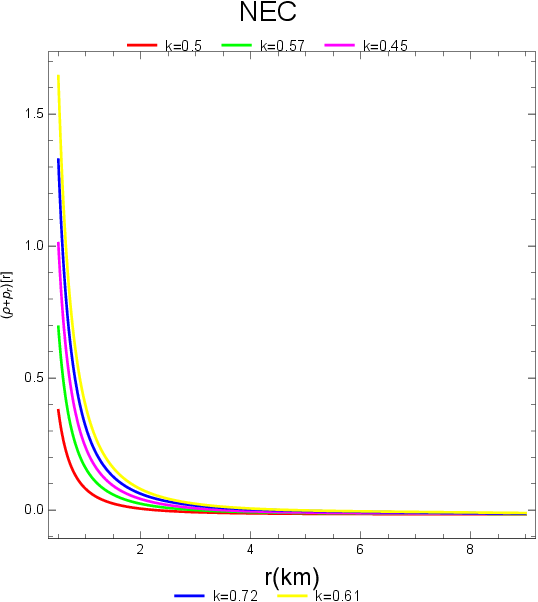}} 
    \caption{(a) $\rho + p_{t}$ with respect to $r$ for parametric perturbation $k=0.5, 0.57, 0.45, 0.72, 0.61$ respectively. \\
    (b)  $\rho + p_{r}$ with respect to $r$ for parametric perturbation $k=0.5, 0.57, 0.45, 0.72, 0.61$ respectively. \\
     (c) $\rho + p_{r} + 2 p_{t}$ with respect to $r$ for parametric perturbation $k=0.5, 0.57, 0.45, 0.72, 0.61$ respectively.}
    \label{fig:energy}
   \end{center}
\end{figure}
Fig \ref{fig:energy} demonstrates the energy conditions such as NEC and SEC with respect to radius for perturbation of model parameter $k$. Here, we can see that given the value of $k$ we have selected, the energy conditions are maintained within the compact stars. This demonstrates the plausibility of the proposed $f(T)$ gravity model.

By confirming the NEC, it is possible to deduce that this energy requirement implies an observer crossing a null diagram, which characterizes the normal matter density as non-negative. The tidal tensor trace that the relating observers test consistently shows a favorable result, according to SEC. Also, it may be said that physical dark matter makes up our solution for describing matter configuration. Thus, by meeting all the necessary requirements, it is possible to conclude that every physical parameter of our given solutions is physically well-behaved.

\subsection{Equation of State parameter}

We have also investigated the possibility of charged matter and whether the dark matter may be deduced from some restrictions that are applicable to Equation of State (EoS) parameter components like $\omega_r$ and $\omega_t$ \cite{bib22,bib23,bib24,bib25}. The transverse pressure and matter density relationship is still unknown, but the radial pressure and matter density are assumed to be linearly related in the model by solving the field equations.
The relationship between pressure and matter density is described by two dimensionless quantities called the equation of state parameters, denoted by $\omega_r$ and $\omega_t$. 
A simplified form of the EoS parameter for our study is
\begin{equation}
\omega _r=\frac{p_r}{\rho}=-\frac{k r^2-2 k \alpha +32 a \pi  r \alpha {Sin}(k r)}{k r^2 y+16 a \pi  r \alpha {Sin}(k r)},
\label{E45}
\end{equation}
\begin{multline}
    \omega _t=\frac{p_t}{\rho}=\frac{1}{k^2 r^2 \left(k^3 r+8 a k \pi  r {Cos}(k r)-8 a \pi  {Sin}(k r)\right) (k r y+16 a \pi  \alpha {Sin}(kr))}\biggl(2 \alpha \biggl( k^6 r^2+16 a^2 \pi ^2 \\
    \left(1+5 k^2 r^2-3 k^4 r^4\right)  +8 a \pi  \biggl( -2 a \pi  \left(1+3 k^2 r^2+k^4 r^4\right) {Cos}(2 k r) -k^3 r \left(2+k^2 r^2\right) {Sin}(k r) \\
    -k r{Cos}(k r)  \biggl( k^3 r \left(-2+k^2 r^2\right) +8 a \pi  \left(1+k^2 r^2\right) {Sin}(k r) \biggr)  \biggr) \biggr) \biggr) .
\label{E46}
\end{multline}
Graphical representations of $\omega_r$ and $\omega_t$ from Eq. (\ref{E45}) and Eq. (\ref{E46}) respectively are
\begin{figure}
   \begin{center}
       \subfigure[]{\includegraphics[width=0.4\textwidth]{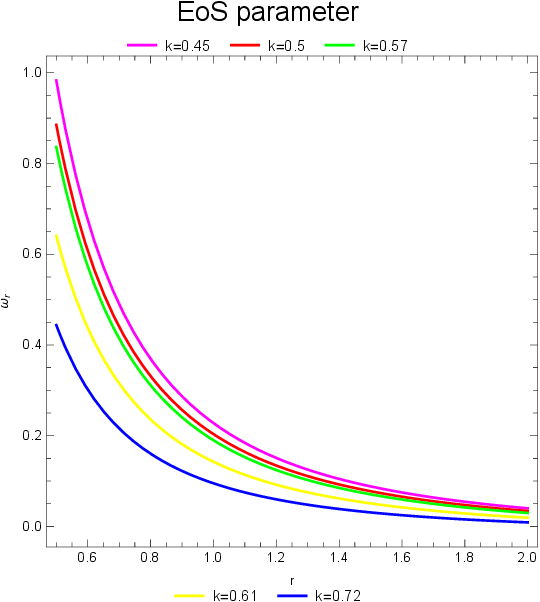}} 
    \subfigure[]{\includegraphics[width=0.4\textwidth]{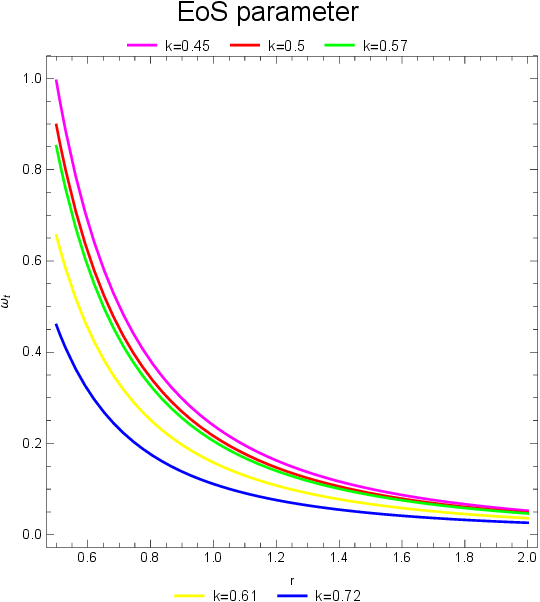}} 
    \caption{(a) EoS parameter of compact star in radial direction from Eq. (\ref{E45}) with respect to $r$ for parametric perturbation $k=0.5, 0.57, 0.45, 0.72, 0.61$ respectively. \\
    (b) EoS parameter of compact star in radial direction from Eq. (\ref{E46}) with respect to $r$ for parametric perturbation $k=0.5, 0.57, 0.45, 0.72, 0.61$ respectively. }
    \label{fig:omega}
   \end{center}
\end{figure}
The right and left panels of Fig \ref{fig:omega} show the graphical representation of the EoS parameter in radial and transverse direction for perturbation of $k$. Anisotropic charged fluids are not exotic in any sense because they have a positive equation of state that is contained inside $0< \omega_r, \omega_t < 1$. At the surface of the star, the Eos parameter disappears and falls monotonically \cite{bib26}.

\subsection{Mass-radius relation}

We generally derive the model mass of the compact objects by using the formula
\begin{equation}
m=\int _0^r4 \pi  x^2\rho (x)dx
\label{E42}
\end{equation}
Additionally, the mass-radius ratio can reveal a pulsar's compactness and thus its precise nature. Buchdahl's limit indicates the maximum mass radius ratio for the isotropic matter distribution, which may be computed using the formula $u=m/r$.  According the mass-radius ratio theory for the stellar objects is categorised as follows:\\
$(i)$ Regular Star :($u \sim 10^{-5}$),
$(ii)$ White Dwarf : ($u \sim 10^{-3}$),
$(iii)$ Neutron Star : ($0.1<u<0.25$),
$(iv)$ Ultra Compact Object : ($0.25<u<0.5$),
$(v)$ Black Hole : ($u=0.5$) \cite{bib20}. 
The mass of the compact objects is
\begin{equation}
m=\frac{r^3 \beta}{12}+\frac{a \pi  r^3 {Sin}(k r)}{k} ,
\label{E43}
\end{equation}
and the mass-radius ratio is
\begin{equation}
u=\frac{m}{r}=\frac{1}{12} r^2 \left(\beta+\frac{12 a \pi  {Sin}(k r)}{k}\right) .
\label{E44}
\end{equation}
The numerical computation for mass and mass-radius ratio have been evaluated in Table \ref{Table:1}, where we can observe for perturbation of $k$ we are getting mass of realistic compact objects such as LMC X-4, Her X-1, 4U 1538-52, SAX J1808.4-3658, and Cen X-3 using Eq. (\ref{E43}). As a result, we may use Eq. (\ref{E44}) to assess the compactness of these objects and determine their nature. In order to verify the stability of the model, the compactness must fall inside $u < 4/9$, which is Buchdahl's limit \cite{bib21}. 
 \begin{table}[!h]
 \begin{center}
     \begin{tabular}{||c |c |c |c |c |c |c |c||} 
 \hline
 Sl. & Star Model & Estimated & Estimated Mass  & $k$ & Model & Compactness & Surface \\[0.5ex] 
 No.  &    & Radius & $M(M_{\odot})$ &  & Mass  &   &  Redshift  \\
  &   &  ($km$) &   &   &  $M(M_{\odot})$   &    &  \\ 
 \hline\hline
 1 & LMC X-4 & 8.301 $\pm$ 0.2 & 1.04 $\pm$ 0.09 &  0.61 & 0.98 & 0.1576 & 0.0895 \\ 
 \hline
 2 & Her X-1 & 8.1 $\pm$ 0.41 & 0.85 $\pm$ 0.15 &  0.45 & 0.95 & 0.11177 & 0.06105 \\ 
 \hline
 3 & 4U 1538-52 & 7.866 $\pm$ 0.21 & 0.87 $\pm$ 0.07 &  0.72 & 0.94 & 0.12533 & 0.06925 \\ 
 \hline
 4 & SAX J1808.4-3658 & 7.951 $\pm$ 1.0 & 0.9 $\pm$ 0.3 &  0.57 & 0.95 & 0.1186 & 0.0625 \\ 
 \hline
  5 & Cen X-3 & 9.178 $\pm$ 0.13 & 1.49 $\pm$ 0.08 &  0.5 & 1.45 & 0.1576 & 0.0895 \\[1ex] 
 \hline\hline
\end{tabular}
\caption{\label{Table:1} Evaluated physical attributes of the compact objects for their estimated masses and radii and parameter $k$}
 \end{center}
\end{table}

\begin{figure}
   \begin{center}
       \subfigure[]{\includegraphics[width=0.4\textwidth]{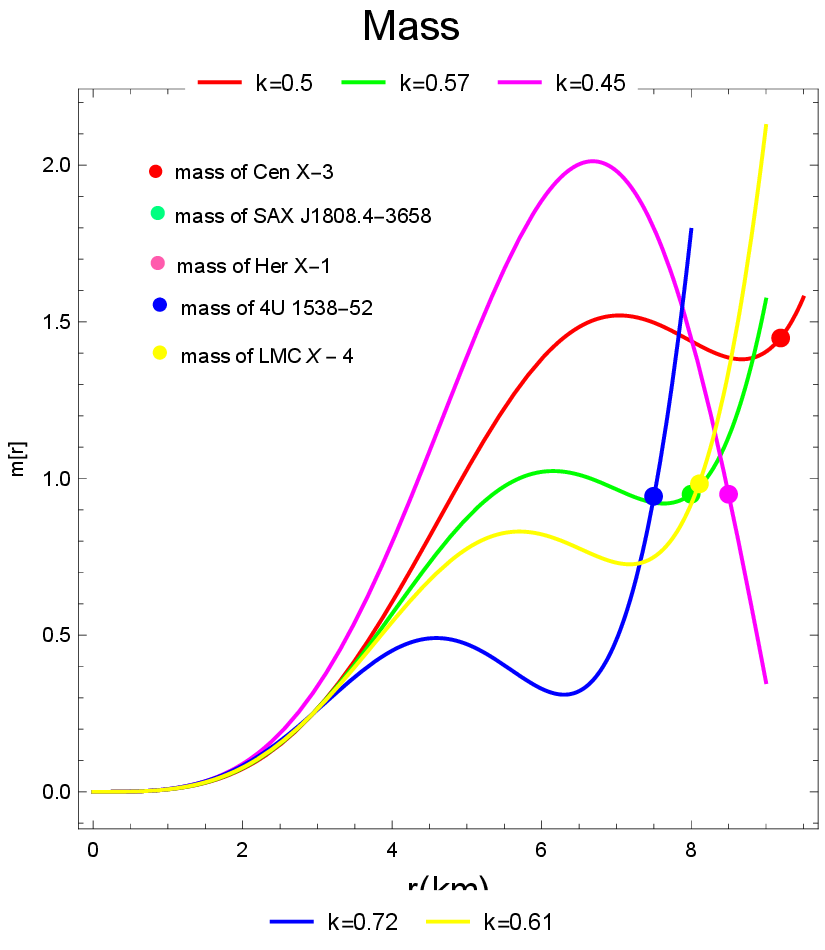}} 
    \subfigure[]{\includegraphics[width=0.4\textwidth]{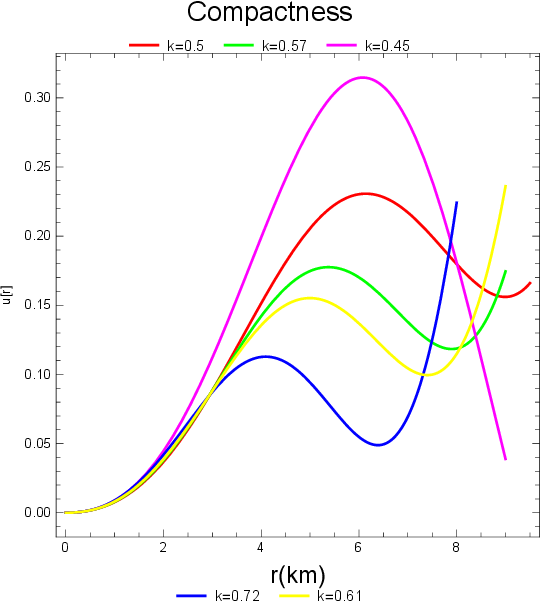}} 
    \caption{(a) Mass of compact star from Eq. (\ref{E43}) with respect to $r$ for parametric perturbation $k=0.5, 0.57, 0.45, 0.72, 0.61$ respectively. \\
    (b) Compactness of compact stars from Eq. (\ref{E44}) with respect to $r$ for parametric perturbation $k=0.5, 0.57, 0.45, 0.72, 0.61$ respectively. }
    \label{fig:mass}
   \end{center}
\end{figure}
The mass and compactness of the aforementioned compact items are shown in Fig. \ref{fig:mass}. The graphic representation in right panel of the figure verifies that our model is stable because it falls inside Buchdahl's limit, which is $u$ smaller than 4/9. Additionally, we may infer from Table \ref{Table:1} that the compact objects indicated above are neutron stars because $u$ falls within (0.1,0.25). Furthermore, the model-generated masses of the neutron star in Table \ref{Table:1} have been evaluated using Eq. (\ref{E43}), and they are within the permissible range of estimated mass.

\section{Equilibrium and stability analysis}

The stability of a compact star model can be verified using various methods. Among them  $(i)$ the causality condition and $(ii)$ the adiabatic index has a great impact.

\subsection{Stability Factor}

Making use of the "cracking" method, which deals with the stability of anisotropic stars under minute radial perturbations According to Herrera \cite{bib27} and Abreu et al. \cite{bib28}, the regions of an anisotropic star that are potentially stable are those where the radial speed of sound crosses the transverse speed of sound. They found that changes in density alone do not cause the system to lose equilibrium for configurations of anisotropic matter. Only perturbations of both density and local anisotropy can cause such deviations. The expansion, collapse, overturning, and cracking were the results of integrating Einstein's equation. Therefore, $0<v^2 _{sr},v^2 _{st}< 1$ is the acceptable limit for the square speed of sound of sound, where $v^2 _{sr} = d{p_r}/d{\rho }$ and $v^2 _{st} = d{p_t}/d{\rho }$. Herrera \cite{bib27} developed the Cracking method in 1922 to assess the stability of the anisotropic matter spheres under radial perturbations. Later, bearing in mind the Cracking approach, Abreu et al. \cite{bib28} investigated the stability of the anisotropic fluid spheres using the stability factor $\mathcal{X}(r)= v^2 _{st} - v^2 _{sr}$.
According to   Abreu et al. \cite{bib28} ,

\begin{math}
    \biggl\{  
\begin{array}{c}
 -1 \leq  \mathcal{X}(r) \leq 0, stable\\
  0 <  \mathcal{X}(r) < 1, unstable
\end{array}
\end{math}

Hence, the square speed of sound in both radial and tangential direction for our model can be expressed as
\begin{equation}
{v_{sr}}^{2}=\frac{dp_r}{{d\rho }}=-\frac{k+8 a k \pi  r^2 {Cos}(k r)-8 a \pi  r {Sin}(k r)}{4 a k \pi  r^2 {Cos}(k r)-4 a \pi  r {Sin}(k r)},
\label{E47}
\end{equation}
\begin{multline}
    {v_{st}}^{2}=\frac{dp_t}{{d\rho }}=\frac{1}{\left(4 a k^2 \pi  r^2 (k r {Cos}(k r)-{Sin}(k r)) \left(k^3 r+8 a k \pi  r{Cos}(k r)-8 a \pi  {Sin}(k r)\right)^2\right)} \biggl(-96 a^2 k^3 \pi ^2 r-k^9 r^3 \\
    -144 a^2 k^5 \pi ^2 r^3-64 a^2 k^7 \pi ^2 r^5-4 a k \pi  r \left(k^6 r^2 \left(7+k^2 r^2\right)+8 a^2 \pi ^2 \left(9+19 k^2 r^2+15 k^4 r^4\right)\right) {Cos}(k r)-16 a^2 k^3 \pi ^2 r \biggl(-6 \\
    +3 k^2 r^2+2 k^4 r^4\biggr){Cos}(2k r)+. 288 a^3 k \pi ^3 r {Cos}(3k r)+224 a^3 k^3 \pi ^3 r^3 {Cos}(3k r)-32 a^3 k^5 \pi ^3 r^5 {Cos}(3k r)+288 a^3 \pi ^3 {Sin}(k r) \\
    +28 a k^6 \pi  r^2 {Sin}(k r)+576 a^3 k^2 \pi ^3 r^2 {Sin}(k r)+640 a^3 k^4 \pi ^3 r^4 {Sin}(k r)+ 4 a k^{10} \pi  r^6 {Sin}(k r)-96 a^3 k^6 \pi ^3 r^6 {Sin}(k r) \\
    +192 a^2 k^4 \pi ^2 r^2 {Sin}(2k r)+96 a^2 k^6 \pi ^2 r^4 {Sin}(2k
r)+16 a^2 k^8 \pi ^2 r^6 {Sin}(2k r)-96 a^3 \pi ^3 {Sin}(3k r)+192 a^3 k^2 \pi ^3 r^2 {Sin}(3k r) \\
+ 256 a^3 k^4 \pi ^3 r^4 {Sin}(3k r)+32 a^3 k^6 \pi ^3 r^6 {Sin}(3k r)\biggr)
\label{E48}
\end{multline}
\begin{figure}
   \begin{center}
       \subfigure[]{\includegraphics[width=0.3\textwidth]{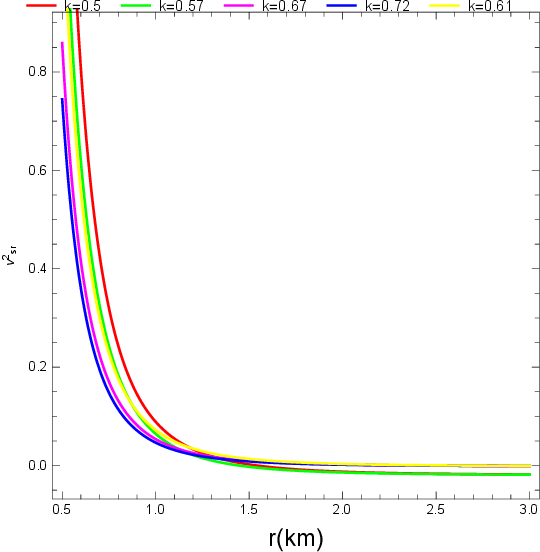}} 
    \subfigure[]{\includegraphics[width=0.3\textwidth]{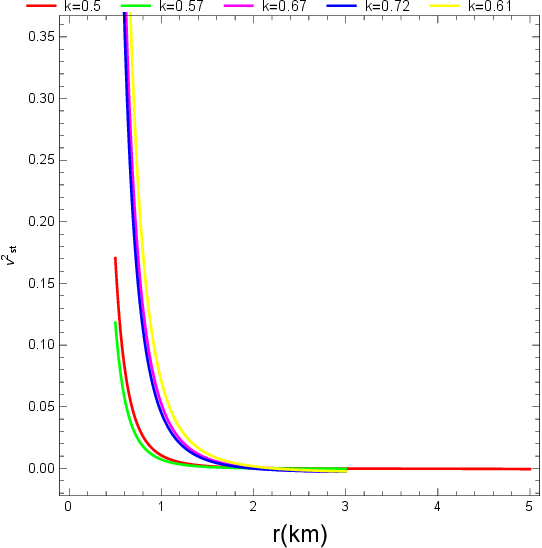}} 
     \subfigure[]{\includegraphics[width=0.3\textwidth]{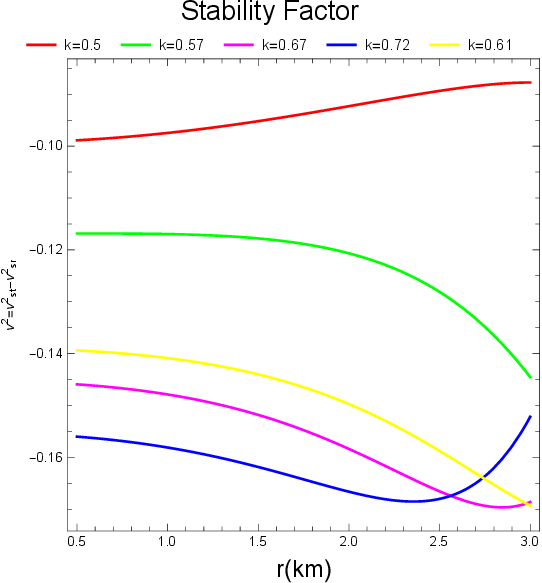}} 
    \caption{(a) Evolution of $v^2 _{sr}$ with respect to $r$ for parametric perturbation $k=0.5, 0.57, 0.45, 0.72, 0.61$ respectively. \\
    (b)   Evolution of $v^2 _{st}$ with respect to $r$ for parametric perturbation $k=0.5, 0.57, 0.45, 0.72, 0.61$ respectively. \\
     (c)  Evolution of $ v^2 _{st} - v^2 _{sr}$ with respect to $r$ for parametric perturbation $k=0.5, 0.57, 0.45, 0.72, 0.61$ respectively.}
    \label{fig:stability}
   \end{center}
\end{figure}
The left and the middle panel figures of Fig. \ref{fig:stability} illustrates the graphical representation of the squared speed of sound in the radial and tangential direction, which confirms that  $0<v^2 _{sr},v^2 _{st}< 1$ within the interior of the compact objects for perturbation of $k$. The right panel figure demonstrates the stability factor introduced by Abreu et al., which confirms that our model is stable as $\mathcal{X}(r)= v^2 _{st} - v^2 _{sr}$ lies with [-1,0].

\subsection{Adiabatic Index}

For a given density, the rigidity of the equation of state is represented by the ratio of the two temperatures or the adiabatic index $\Gamma =\left(\frac{\rho +p_r}{p_r}\right)\frac{dp_r}{{d\rho }}$. Under an infinitesimal radial adiabatic effect, this adiabatic index $\Gamma$ can be utilized to study the dynamical stability of the star structure; it should be bigger than 4/3 in the interior area \cite{bib30}. This limiting number was determined by Chandrashekhar to assess the dynamical stability of an isotropic relativistic star with spherical symmetry in the presence of a minor radial adiabatic perturbation \cite{bib29}. However, $\Gamma (r) > 4/3$ inside $r < R$ is required for anisotropy, or anisotropic matter structure, stability \cite{bib31,bib32}. 
\begin{equation}
\Gamma =\frac{\left(k+8 a k \pi  r^2 {Cos}(k r)-8 a \pi  r {Sin}(k r)\right) \left(k \left(2 \alpha+r^2 (-1+y)\right)-16 a \pi  r \alpha {Sin}(k
r)\right)}{4 a \pi  r (k r {Cos}(k r)-{Sin}(k r)) \left(k \left(r^2-2 \alpha \right)+32 a \pi  r \alpha {Sin}(k r)\right)}
    \label{E49}
\end{equation}

\subsection{Surface Redshift}

In physics and general relativity, the phenomenon of electromagnetic waves or photons leaving a gravitational well is known as gravitational redshift (also known as the Einstein shift in earlier literature) \cite{bib33}. A redshift is the consequence of this energy loss and is typically defined as a decrease in wave frequency and an increase in wavelength. For compatibility, the surface redshift is determined in ($z_{s}$) $\leq$ 5.211 \cite{bib34}. The surface redshift is generally derived from $z_s=(1-2u)^{-1/2} -1$. Surface redshift for our model derived as
\begin{equation}
z=-1+\frac{1}{\sqrt{1-\frac{1}{12} r^2 \left(\beta+\frac{12 a \pi  {Sin}(k r)}{k}\right)}}
    \label{E50}
\end{equation}

\begin{figure}
   \begin{center}
       \subfigure[]{\includegraphics[width=0.4\textwidth]{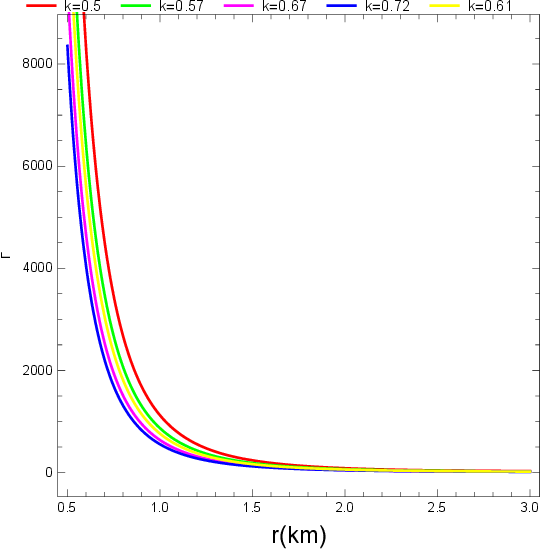}} 
    \subfigure[]{\includegraphics[width=0.4\textwidth]{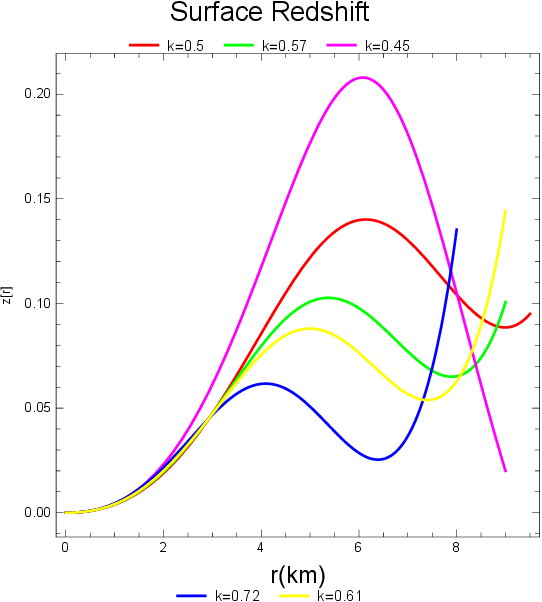}} 
    \caption{(a) Adiabatic index of compact star from Eq. (\ref{E49}) with respect to $r$ for parametric perturbation $k=0.5, 0.57, 0.45, 0.72, 0.61$ respectively. \\
    (b) Surface Redshift of compact stars from Eq. (\ref{E50}) with respect to $r$ for parametric perturbation $k=0.5, 0.57, 0.45, 0.72, 0.61$ respectively. }
    \label{fig:adiabatic}
   \end{center}
\end{figure}
The right panel figure in Fig \ref{fig:adiabatic} shows the evolution of the adiabatic index, which is bigger than 4/3 in the interior of the compact objects and the left panel demonstrates the evolution of surface redshift, which is less than 5.211 within the compact objects. Hence, we can conclude that our model is well stable for the perturbation of $k=0.5, 0.57, 0.45, 0.72, 0.61$.

\section{Moment of Inertia}

According to Lattimer et. al. \cite{bib40} the moment of inertia for a uniformly spinning neutron star is
\begin{equation}
I=\frac{8 \pi }{3}\int _0^Rr^4\left(\rho +p_r\right)e^{(\lambda -\nu )/2}. \frac{\bar{\omega }}{\Omega }dr ,
    \label{E66}
\end{equation}
where $\Omega$ is the angular velocity and $\bar\omega$ is the rotational drag. Hartle's equation \cite{bib41} satisfies the rotational drag as
\begin{equation}
    \frac{d}{{dr}}\left(r^4j\frac{d\bar{\omega }}{{dr}}\right)=-4r^3\bar{\omega } \frac{{dj}}{{dr}} ,
    \label{E67}
\end{equation}
for $j=e^{-(\lambda +\nu )/2}$ at $j(R)=1$. Another proposed form of inertia is by Bejger et. al. \cite{bib42} for a non-rotating neutron star is
\begin{equation}
    I=\frac{2}{5}(1+x)M R^2 ,
    \label{E68}
\end{equation}
where $x=\frac{M}{R}$ $M_{\odot}/km$. Hence Eq. (\ref{E68}) reformulated for our model as
\begin{equation}
I=\frac{r^5 (k y+12 a \pi {Sin}(k r)) \left(k \left(12+r^2 y\right)+12 a \pi  r^2 {Sin}(k r)\right)}{360 k^2}
    \label{E69}
\end{equation}
\begin{figure}
   \begin{center}
       \subfigure[]{\includegraphics[width=0.4\textwidth]{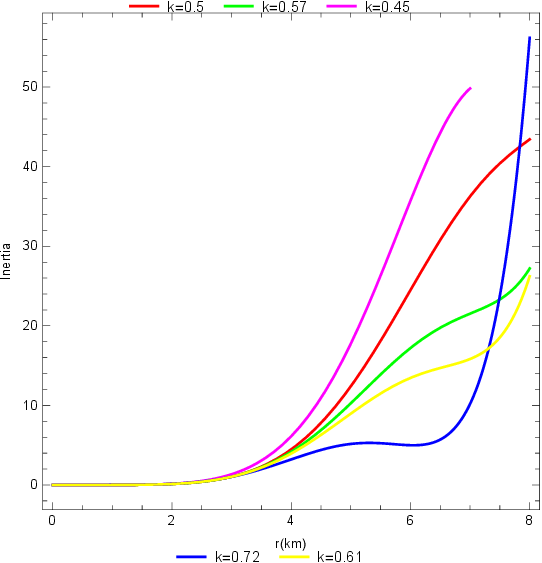}} 
    \subfigure[]{\includegraphics[width=0.4\textwidth]{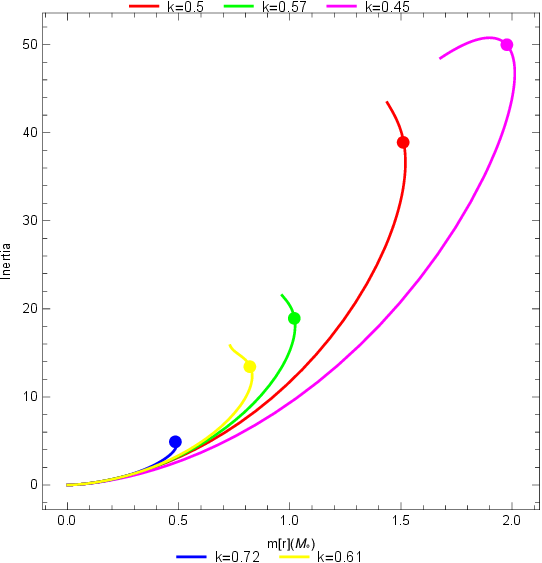}} 
    \caption{(a) Moment of Inertia of compact stars from Eq. (\ref{E69}) with respect to $r$ for parametric perturbation $k=0.5, 0.57, 0.45, 0.72, 0.61$ respectively. \\
    (b) Moment of Inertia of compact stars from Eq. (\ref{E69}) with respect to mass ($m$) for parametric perturbation $k=0.5, 0.57, 0.45, 0.72, 0.61$ respectively. }
    \label{fig:inertia}
   \end{center}
\end{figure}
Visual illustration of Eq. (\ref{E69}) in the left and right panels of Fig. \ref{fig:inertia} show the moment of inertia with respect to mass and radius, respectively, and show that for the compact stars \emph{LMC X-4}, \emph{Her X-1}, \emph{4U 1538-52}, \emph{SAX J1808.4-3658}, and \emph{Cen X-3}, the moment of inertia increases monotonically with increasing mass. It can be seen that in the figure on the left panel, the maximum moment of inertia is represented by blue, yellow, red, magenta, and red dots. We have established that the moment of inertia increases monotonically with mass, as in \cite{bib1}, and we have also clearly shown that $M_{max}$ is not equal to $I_{max}$.

\section{Discussion and Conclusion}

We have assessed the compact objects in the $f(T)$ gravity theory backdrop in this study, using the \emph{Bose-Einstein DM} density profile as an internal fluid of the compact objects. The $f(T)$ gravity theory is a powerful tool for studying the physical acceptability of compact objects. It was first presented as the Teleparallel Equivalence of General Relativity (TEGR). In TEGR, the theory is constructed using tetrad fields rather than metric tensors. Torsion is the sole impact that persists since the curvature at each point is zero due to the usage of a tangent space frame of reference. As a result, TEGR explains that space-time torsion is what causes gravity. A spherically symmetric static solution in $f(T)$ gravity models was recently studied by Wang et al. \cite{bib35}. It is discovered that the class of $f(T)$ gravity models that can be solved in this frame is limited. The relativistic neutron star solution in $f(T)$ gravity models can only be achieved, according to another study's authors \cite{bib36}, if $f(T)$ is a linear function of the torsion scalar $T$. This work has explored the black hole with a cosmological constant.
Because \emph{Bose-Einstein DM} is extremely dense and composed of neutral hydrogen clouds, we have utilized it as an internal fluid. It is also believed to be a cold bosonic gravitationally confined system. An article contains a systematic investigation of the characteristics of Bose-Einstein Condensed Galactic Dark Matter halos \cite{bib37}. Furthermore, the notion that some types of BEC could arise in neutron stars has long been entertained \cite{bib38}. We list below the basic work done in this paper and our actual findings pointwise.

(i) Using the exterior and interior matching conditions of the Schwarzchild solution, we developed the metric potential functions $e^\eta$ and $e^\zeta$ for \emph{Bose-Einstein DM}. Fig. \ref{fig:metric} shows that metric potentials are non-singular in nature, also $e^{\eta(0)}=1$ and $e^{\zeta(0)}=1$. Since the metric potential is appropriate for studying non-singular compact objects, we thoroughly assess the physical characteristics of the objects. 

(ii) The density, radial pressure, and tangential pressure of the compact objects have been evaluated in Eq. (\ref{E32}), Eq. (\ref{E33}), and Eq. (\ref{E34}) respectively. We have calculated the first and second order derivatives of the density in order to $d\rho /dr =0$, and $d^2 \rho /dr < 0$ and determined the maximum density distribution for $r \rightarrow 0$. i.e., the cores of the compact object possess maximum density. In Figs. \ref{fig:density} and \ref{fig:ani}, density, radial pressure, tangential pressure, and anisotropic factor are illustrated graphically. As we can see, the pattern of density, radial pressure, and tangential pressure is decreasing, while the largest value is found in the vicinity of the center. Additionally, there is asymptotic flatness for both tangential and radial pressure close to the objects' surfaces. Hence, the radial and tangential pressure vanishes at the surface of the compact objects. The anisotropic factor exhibits favorable trade-offs across the compact object region. Density and pressure in both the transverse and radial directions must be understood in order to comprehend the nature of the compact object. There shouldn't be any singularities in the physical parameters of compact objects because their density and pressure should peak around their center and then decrease monotonically over their outermost layer. Moreover, for any compact object, $p_{t} > p_{r}$, the anisotropic factor, represented by $\Delta=p_{t}-p_{r}$, indicates that the pressure is not equal in both the transversal and radial directions. 

(iii) To further validate our results we have evaluated the gradient of density, radial pressure, and tangential pressure, where we can see $d\rho/dr$, $dp_r /dr$, and $dp_t /dr$ have negative values throughout the region in Fig \ref{fig:gradient} for perturbation of $k=0.5, 0.57, 0.45, 0.72, 0.61$. This confirms that our model is non-singular in nature and well-behaved for further study as discussed by the authors in \cite{bib39}.

(iv) Energy conditions need to be assessed in order to validate the singularity and the model. The non-exotic nature of the substance is likewise confirmed by the energy conditions. The graphical alignment of the energy condition is shown in Fig \ref{fig:energy}, confirming that the energy criteria are satisfied for perturbation $k=0.61,0.72,0.57,0.45,0.5$.

(v) The EoS parameter has been verified both in radial and tangential direction and demonstrated in Fig \ref{fig:omega}. The dimentionless quantity $\omega$ helps to establish the relation between pressure and matter density and for being a non-exotic fluid it must be positive and lies within (0,1).

(vi) Masses of the compact objects have been evaluated using Eq. (\ref{E43}). For perturbation of $k$ we get masses of different realistic compact objects such as for $k=0.61$ $\rightarrow$ \emph{LMC X-4}, $k=0.45$ $\rightarrow$ \emph{Her X-1}, $k=0.72$ $\rightarrow$ \emph{4U 1538-52}, $k=0.57$ $\rightarrow$ \emph{SAX J1808.4-3658}, and $k=0.5$ $\rightarrow$ \emph{Cen X-3}. The derived model masses and mass-radius relation values for the aforementioned compact objects are displayed in Table \ref{Table:1}. Fig \ref{fig:contour1} and Fig \ref{fig:contour2} demonstrate the radially symmetric density profile, which confirms that the core of the star has the maximum density and vanishes at the surface as discussed in \cite{bib1,bib18} as well. These objects are classified as \emph{Neutron Star}, with $0.1<u<0.25$, according to the compactness. Furthermore, Buchdahl's limit is reached by the compactness, indicating that the model is stable.

(vii) The stability and causality conditions have also been verified. The square speed of sound in both radial and tangential directions must be within (0,1) for a stable model, and the stability factor $\mathcal{X}$ should lie within [-1,0] for the cracking method. Fig \ref{fig:stability} confirms that our model is well stable for the realistic neutron star \emph{LMC X-4}, \emph{Her X-1}, \emph{4U 1538-52}, \emph{SAX J1808.4-3658}, and \emph{Cen X-3}. We have also verified the adiabatic index and the surface redshift in Fig \ref{fig:adiabatic}, which confirms we have a stable model for \emph{Bose-Einstein Dark Matter} density profile in the backdrop of $f(T)$ gravity theory. 

(viii) When studying non-rotating compact objects, inertia is a crucial factor. The moment of inertia of a non-rotating neutron star must be growing in relation to mass change. The moment of inertia with respect to radius and mass function for the compact objects \emph{LMC X-4}, \emph{Her X-1}, \emph{4U 1538-52}, \emph{SAX J1808.4-3658}, and \emph{Cen X-3} is visually illustrated in Fig. \ref{fig:inertia}. The data indicates that while maximum mass value is not equal to maximum inertia for the perturbation of $k=0.61,0.72,0.57,0.45,0.5$, inertia is increasing monotonically.

\section*{Acknowledgments}

P.R. acknowledges the Inter-University Centre for Astronomy and Astrophysics (IUCAA), Pune, India for granting visiting associateship.

\section*{Data Availability Statement}

No data was generated or analyzed in this study.

\section*{Conflict of Interest}

There are no conflicts of interest.

\section*{Funding Statement}

There is no funding to report for this article.


\end{document}